\documentclass[aps,onecolumn,showpacs]{revtex4}
\usepackage{graphicx,color}
\usepackage{amsfonts}
\usepackage{amsmath}
\usepackage{amssymb}
\usepackage[T1]{fontenc}
\usepackage[portuges]{babel}

\newcommand{\bq}{\begin{eqnarray}}
\newcommand{\eq}{\end{eqnarray}}
\newcommand{\be}{\begin{equation}}
\newcommand{\ee}{\end{equation}}

\newcommand{\tr}{\operatorname{tr}}

\definecolor{gray}{rgb}{.4,.4,.4}
\definecolor{deepgreen}{rgb}{.1,.6,.3}

\begin{document}

\title{Mixedness and entanglement for two-mode Gaussian states}
\author{L.~A.~M.~Souza$^{1}$}
\author{R.~C.~Drumond$^2$}
\author{M.~C.~Nemes$^2$}
\author{K. M. Fonseca Romero$^3$}
\affiliation{$^1$ \emph{Campus} Florestal - Universidade Federal de Vi\c{c}osa, LMG 818 -
Km 6, CEP 35.960-000, Florestal, Minas Gerais, Brazil.}
\affiliation{$^2$Departamento de F\'{\i}sica, Instituto de Ci\^{e}ncias Exatas,
Universidade Federal de Minas Gerais, CP 702, CEP 30123-970, Belo Horizonte,
Minas Gerais, Brazil.}
\affiliation{$^3$Universidad Nacional de Colombia - Bogot\'a, Facultad de Ciencias,
Departamento de F{\'isica}, Grupo de {\'O}ptica e Informaci{\'o}n Cu{\'a}%
ntica, Carrera 30 Calle 45-03, C.P. 111321, Bogot{\'a}, Colombia}

\begin{abstract}
We analytically exploit the two-mode Gaussian states nonunitary
dynamics. We show that in the zero temperature limit, entanglement
sudden death (ESD) will always occur for symmetric states (where
initial single mode compression is $z_0$) provided the two mode
squeezing $r_0$ satisfies $0 < r_0 < \frac{1}{2} \log (\cosh (2
z_0)).$ We also give the analytical expressions for the time of ESD.
Finally, we show the relation between the single modes initial
impurities and the initial entanglement, where we exhibit that the
later is suppressed by the former.
\end{abstract}

\maketitle

\section{Introduction}

The concept of entanglement, a typically quantum mechanical
property, is a natural consequence of the superposition principle
for composite systems. It was discussed by Schr\"{o}dinger who
immediately realized its (at the time) seemingly ``unphysical''
consequences \cite{schrodinger}. Nowadays the growing interest in
the subject is related to quantum information theory and potential
technological applications \cite{esd, Bacon-Phys}. Thus a lot of
effort has been put both in the quantification of entanglement and
in studying the consequences of deleterious environmental effects on
this property \cite{decoherence}. It is well known today, both
theoretically and experimentally, that in general such effects tend
to destroy quantum properties; in the case of systems with one
degree of freedom this happens only asymptotically \cite{leocarol1,
leocarol2, adesso1}. The studies concerning the degradation of
quantum effects are vast in the literature, specially those related
to continuous variables systems subjected to noisy channels, see
\cite{cite1, cite2}. Some recent works devote their attention to the
robustness of Gaussian and non-Gaussian states under dissipative
channels \cite{cite3}.

It was recently realized that in bipartite systems that entanglement
may disappear suddenly, phenomenon called therefore ``entanglement
sudden death'' (ESD) or finite-time disentanglement \cite{esdcv,
PR}. The phenomenon is strongly related to geometrical properties of
the set of mixed quantum states \cite{GeoSudDeath}, but from a
physical point of view, the issue is still far from being closed.

In the present work we study more specifically the entanglement properties
of two mode Gaussian states under a nonunitary evolution. In ref. \cite%
{leocarol1, leocarol2, adesso1} the entropy growth of single-mode
Gaussian and non-Gaussian states coupled to a reservoir has been
presented in analytical form. We show that in the two-mode case,
single-mode squeezing plays an important role in entanglement
dynamics, even for fully symmetric channels (phenomenon that appears
also in the case of qu-bits \cite{marcelin}). For the specific case
of a zero temperature reservoir and symmetric states (where initial
single mode squeezing of both modes is $z_0$) we are able wo show
that ESD will always occur if $0 < r_0 < \frac{1}{2} \log (\cosh (2
z_0))$, where $r_0$ is the two mode squeezing. We also show that
there exists an upper limit for the degree of mixedness of a state
so that it can exhibit entanglement. This result may turn out useful
for experimental realizations of two-mode entangled Gaussian states.
Finally, for symmetric and pure states evolving in a reservoir at
zero temperature, we analytically present the time of ESD.

The paper is organized as follows: in section \ref{gaussiangeneral}
we present some well known properties concerning Gaussian States; in
the next section we briefly review the analytical dynamics of
two-mode Gaussian states; in section \ref{results} we present our
results. Finally, some conclusions are drawn.


\section{General properties of two-mode Gaussian states}

\label{gaussiangeneral}

A two-mode Gaussian state with vanishing averages $\langle{a}_i\rangle=
\tr (a_i \rho)=0$, {$i=1,2$}, can always be written as
\begin{equation}  \label{rhodoismodos}
\rho_G = S_1 (z_1, z_2) S_2 (r) \mathbf{\sigma}(\nu_1, \nu_2) S_2^\dagger(r)
S_1^\dagger (z_1, z_2),
\end{equation}
where $S_1(z_1, z_2)$ is the single mode squeezing operator, $S_2(r)$ is the
two-mode squeezing operator and $\mathbf{\sigma}(\nu_1, \nu_2)$ is the
two-mode thermal state. More explicitly we have:
\begin{eqnarray}
S_1 (z_1, z_2) &=& \exp \left[ \frac{z_1}{2}(a_1^{\dagger 2} - a_1^2) \right]
\exp \left[ \frac{z_2}{2}(a_2^{\dagger 2} - a_2^2) \right], \\
S_2 (r) &=& \exp \left[ r (a_1^\dagger a_2^\dagger - a_1 a_2) \right], \\
\mathbf{\sigma}(\nu_1, \nu_2) &=& \frac{1}{\nu_1+1} \frac{1}{\nu_2 + 1}
\sum_n \sum_m \left( \frac{\nu_1}{\nu_1 + 1} \right)^n \left( \frac{\nu_2}{%
\nu_2 + 1} \right)^m\left| n \right\rangle \left\langle n \right| \otimes
\left| m \right\rangle \left\langle m \right|,  \notag
\end{eqnarray}
where $z_i$ is the single-mode squeezing parameter of the mode $i$, $%
a_i^{(\dagger)}$ is the annihilation (creation) operator of the $i-$th mode,
$r$ is the two-mode squeezing and $\nu_i$ is the ``mixedness'' of the $i-$th
mode (by ``mixedness'' we mean the initial number of thermal photons of the
state), related of the thermal two-mode state $\mathbf{\sigma}$. We assume $%
\langle a_i \rangle = 0$ in this work, since the entanglement properties are
independent of them, and in our equations of motion the second momenta
evolution decouple form the first momenta. We also choose, without loss of
generality in this case, the squeezing parameters $z_i$ and $r$ to be real.

This state is entirely described by the parameters given above. However, in
order to handle entanglement properties, it is most convenient to write
these parameters in terms of the corresponding covariance matrix (CM):
\begin{eqnarray}  \label{MC}
V_{{\rho}}=\left(
\begin{array}{cccc}
n_{1}+\frac{1}{2} & m_{1} & m_{s} & m_{c} \\
m_{1}^{*} & n_{1}+\frac{1}{2} & m_{c}^{*} & m_{s}^{*} \\
m_{s}^{*} & m_{c} & n_{2}+\frac{1}{2} & m_{2} \\
m_{c}^{*} & m_{s} & m_{2}^{*} & n_{2}+\frac{1}{2}%
\end{array}
\right),
\end{eqnarray}
where $n_{i}=\langle {a}^{\dagger}_{i}{a}_{i}\rangle$, $%
m_{i}=-\langle {a}_{i}^{2}\rangle$, $m_{s}=-\langle {a}_{1}{a}%
_{2}^{\dagger}\rangle$, $m_{c}=\langle {a}_{1}{a}_{2}\rangle$, and $%
\langle {\xi} \rangle$ denotes the quantum expectation value $\tr%
({\xi}{\rho})$ of an observable ${\xi}$. The CM can also be
written as
\begin{eqnarray}
V_{{\rho}}=\left(
\begin{array}{cc}
V_1 & C \\
C^\dagger & V_2%
\end{array}
\right),
\end{eqnarray}
where $V_i$ is a $2 \times 2$ matrix related to the mode $i$, and $C$ is a $%
2 \times 2$ matrix that gives the correlations (both quantum and classical)
between the modes. For later use, we define some invariants of the
covariance matrix as \cite{adesso1, haruna}:
\begin{eqnarray}
I_V &=& \det V_\rho,  \notag \\
I_{1,2} &=& \det V_{1,2},  \notag \\
I_3 &=& \det C,  \notag \\
I_4 &=& \tr \left[ V_1 Z C Z V_2 Z C^\dagger Z \right].
\end{eqnarray}
These quantities are invariants under local unitary operations and $Z=\text{%
diag}\{1,-1\}$. Next we give the explicit connection between the parameters
of the Gaussian state and the matrix elements of the covariance matrix:
\begin{eqnarray}
z_i&=& \frac{1}{2} \mathrm{arctanh}\left[ \frac{m_i}{n_i + \frac{1}{2}} %
\right] \\
\nu_1 &=& \frac{1}{2} \left( \det V_1 - \det V_2 \right)+  \notag \\
&& + \frac{1}{2} \sqrt{1-x^2} \left( \det V_1 + \det V_2 \right) - \frac{1}{2%
}, \\
\nu_2 &=& \frac{1}{2} \left( \det V_2 - \det V_1 \right)+  \notag \\
&& + \frac{1}{2} \sqrt{1-x^2} \left( \det V_1 + \det V_2 \right) - \frac{1}{2%
}, \\
r &=& \frac{1}{2} \mathrm{arctanh}\left[ x \right],  \label{rt}
\end{eqnarray}
where we have defined
\begin{equation}
x = \frac{2 m_s}{ (\sqrt{\det V_1} + \sqrt{\det V_2}) \sinh (z_1 + z_2)}.
\end{equation}
Once the evolution of the covariance matrix is obtained, the
evolution of the parameters of the state can be inferred.

Since we are working with Gaussian states, there are \emph{necessary
and sufficient} criterion to determine if the state is entangled
\cite{simon, DGCZ, serafini1}. Simon \cite{simon} has shown that for
any two-mode Gaussian state, if the following inequality is observed
\begin{equation}  \label{simonquantity}
S\left( V_{\hat{\rho}} \right) = I_1 I_2 + (1/4-|I_3|)^2 - I_4 - 1/4
(I_1+I_2) \geq 0,
\end{equation}
the state is separable. For the purposes of this work, Simon's
criterion is enough to study the entanglement dynamics.

\section{Nonunitary dynamics and its analytical solution}

The degradation of the quantum information content of Gaussian states is a
subject of interest, both for the technological and/or experimental
applications as well as for fundamental quantum mechanics in what concerns
the classical limit.

The usual approach to non-unitary dynamics is by means of master equations,
which has found several successful applications in other areas of physics.
Our master equation reads
\begin{equation}  \label{mestra}
\dot{\rho} = \mathcal{L} \rho,
\end{equation}
where $\mathcal{L}$ is a superoperator given by:
\begin{eqnarray}  \label{liouvilliano1}
\mathcal{L} &=& \sum_{i = 1,2} \left(\gamma_i (n_{i}^B + 1) (2 a_i \bullet
a_i^\dagger - a_i^\dagger a_i \bullet - \bullet a_i^\dagger a_i) + \gamma_i
~ n_i^B (2 a_i^\dagger \bullet a_i - a_i a_i^\dagger \bullet - \bullet a_i
a_i^\dagger)\right).
\end{eqnarray}
In the equation above, 
$\gamma_i$ is the dissipation constant of the reservoir related to the mode $%
i$, $\bar{n}_i^B$ is related to the temperature of the thermal bath of the
mode $i$, and $a_i^\dagger (a_i)$ is the creation (annihilation) operator of
the respective mode.

The equation \eqref{liouvilliano1} models a linear coupling between the
state (the two-mode Gaussian state in our case) with a thermal bath of
quantum harmonic oscillators. The evolution of each term of the covariance
matrix, evolving under the dynamics described above, is given by:
\begin{eqnarray}
n_i &=& \mathrm{e}^{-2 t \gamma_i} \left(\left(-1+e^{2 t \gamma_i}\right)
n_i^B + \cosh (2 z_i^0) \left( \nu_i^0 \cosh^2 r_0 + (1 + \nu_k^0) \sinh^2
r_0 \right)+\sinh^2 z_i^0 \right); \\
m_i &=& - \mathrm{e}^{-2 t \gamma_i} (\nu_i^0 - \nu_k^0 +(1 + \nu_i^0 +
\nu_k^0) \cosh (2 r_0)) \cosh z_i^0 \sinh z_i^0; \\
m_c &=& \frac{1}{2} \mathrm{e}^{-t (\gamma_1 + \gamma_2)} (1 + \nu_1^0 +
\nu_2^0) \cosh (z_1^0 + z_2^0) \sinh (2 r_0); \\
m_s &=& - \frac{1}{2} \mathrm{e}^{-t (\gamma_1 + \gamma_2)} (1 + \nu_1^0 +
\nu_2^0) \sinh (2 r_0) \sinh(z_1^0 + z_2^0).
\end{eqnarray}
In the equations above, $i = 1$ or $2$ (mode 1 or mode 2) and $k \neq i$.
The parameters are such that 
$z_i^0$ is the initial \emph{single mode squeezing}, $\gamma_i$ is the
dissipation constant, $\nu_i^0$ is related to the \emph{initial mixedness}
and $n_i^B$ is the reservoir temperature, where all the quantities refer to
the $i-$th mode, as denoted by the suscript $i$. Also, we have that $r_0$ is
the initial \emph{two-mode squeezing}.

\section{Results}

\label{results}

\subsection{Entanglement dynamics and single mode squeezing}

In order to get a clear picture and to gain physical insight into
the rich and multifaceted aspects of the non-unitary dynamics of
general two-mode Gaussian states, we consider firstly the simplest
case, \emph{i.e.} the two-mode squeezed vacuum in dissipative
channel.

\begin{figure}[!h]
\begin{center}
\vspace*{0cm} 
\includegraphics[bb = 0 0 300 182]{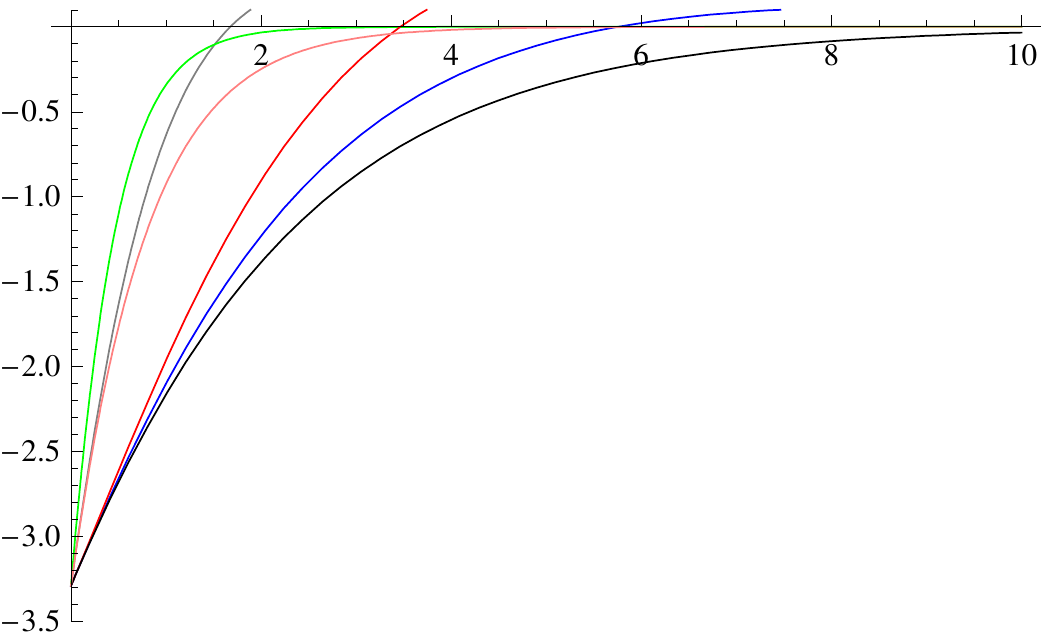}
\end{center}
\caption{Simon criterion for several cases of initial parameters.
The common parameters are: $z_1^0 = z_2^0 = \protect\nu_{1}^0 =
\protect\nu_{2}^0 = 0,
r_0 = 1$. In each curve: gray: $\protect\gamma_1 = 0.1, \protect%
\gamma_2 = 0.5, n_1 = 0.2, n_2 = 0.2$; blue: $\protect\gamma_1 = 0.1,
\protect\gamma_2 = 0.1, n_1 = 0.2, n_2 = 0.2$; green: $\protect\gamma_1 =
0.5, \protect\gamma_2 = 0.5, n_1 = 0.0, n_2 = 0.0$; red: $\protect\gamma_1 =
0.1, \protect\gamma_2 = 0.1, n_1 = 1.0, n_2 = 0.0$; black: $\protect\gamma_1
= 0.1, \protect\gamma_2 = 0.1, n_1 = 0.0, n_2 = 0.0$; pink: $\protect\gamma%
_1 = 0.1, \protect\gamma_2 = 0.5, n_1 = 0.0, n_2 = 0.0$.}
\label{figura1}
\end{figure}

Let us consider the case of a two-mode vacuum state without
single-mode squeezing (\emph{i.e.} $z_{1}=z_{2}=0$). As shown in
Figure \ref{figura1}, for symmetric and asymmetric channels,
\emph{with temperatures $n_{1}^{B}$ and $n_{2}^{B}$ different from
zero}, there will always be a finite time when entanglement
vanishes. This can be understood using a geometrical picture of
entanglement decay \cite{GeoSudDeath}. In fact, in this case, the
long-time state is a separable mixed state, well within the set of
separable states. Thus, if an initial state is entangled, it
necessarily crosses the border of separable states in finite time.
For zero-temperature baths, i.e. if $n_{1}^{B}=n_{2}^{B}=0$, even if
one uses different dissipation constants $\gamma _{1}$ and $\gamma
_{2}$,
 the entanglement only disappears asymptotically.

\begin{figure}[!h]
\begin{center}
\vspace*{0cm} 
\includegraphics[bb = 0 0 300 192]{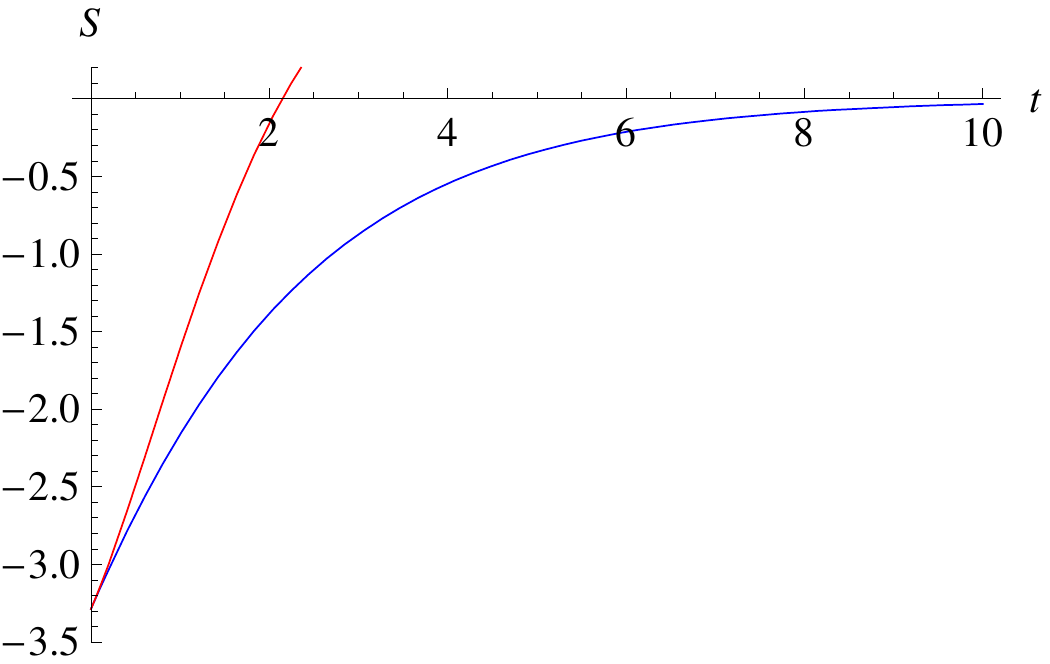}
\end{center}
\caption{Simon criterion for two set of initial parameters, say: $\protect\nu%
_{1}^0 = \protect\nu_{2}^0 = 0, r_0 = 1$. Blue: $\protect\gamma_1 = \protect%
\gamma_2 = 0.1, n_{1}^B = n_{2}^B = 0, z_1^0 = z_2^0 = 0$; Red: $\protect%
\gamma_1 = \protect\gamma_2 = 0.1, n_{1}^B = n_{2}^B = 0, z_1^0 = 2, z_2^0 =
0 .$ }
\label{figura2}
\end{figure}

Next we introduce single mode squeezing, \emph{i.e.} $z_1^0 \neq 0$ and/or $%
z_2^0 \neq 0$. We note in Figure \ref{figura2} that, even for the zero
temperature case, one observes entanglement sudden death (ESD). We note that
there is a dynamical entropy increase of the squeezed mode, caused by the
reservoir, which acts in such a way that, for a relatively short time
interval, this mode's entropy grows and then decays to the vacuum. We have
observed that this dynamical entropy growth \cite{leocarol1, leocarol2}
causes the 
entanglement disappearance in finite time.

\begin{figure}[!htp]
\begin{center}
\vspace*{0cm} 
\includegraphics[bb = 14 14 315 172]{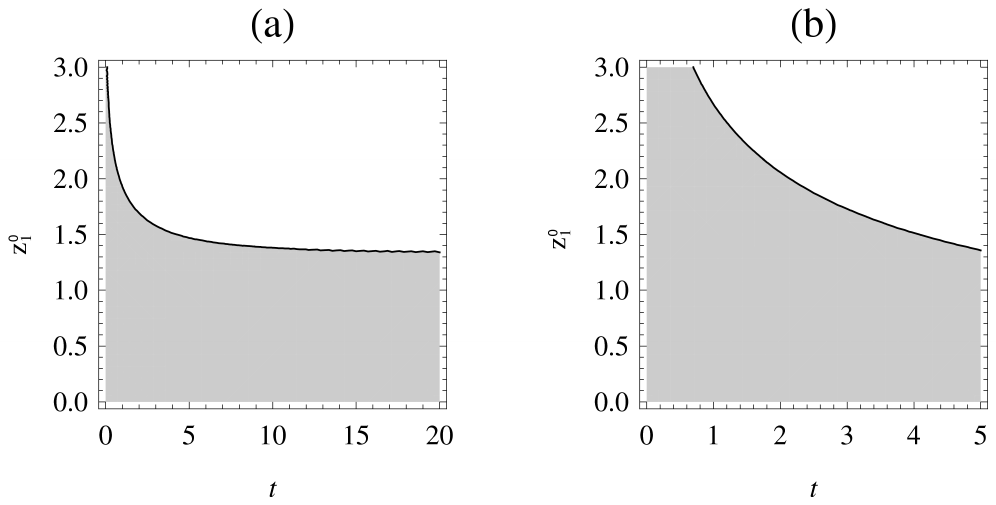}
\end{center}
\caption{Simon criterion for values of the initial parameter $z_1^0$
and time $t$, where we have used $z_1^0 = z_2^0$. For the shaded
area the Simon criterion $S$ is negative (entangled state), in the
white area $S$ is positive. Note that
there is a limit for $z_1$ in which the state will have ESD. In this case: $%
z_1 \simeq 1.4$. Parameters: $r_0 = 1, \protect\nu_{1}^0 = \protect\nu_{2}^0
= 0, n_{1}^B = n_{2}^B = 0, \protect\gamma_1 = \protect\gamma_2 = 0.1.$
Graphs (a) and (b) are the same function, in a different time scale.}
\label{figura7}
\end{figure}

Since single-mode squeezing turns out to play a significant role on
 ESD. In Figure \ref{figura7} we show Simon's criterion evolution for states
 with compression in only one of the modes, for reservoirs at null temperature.
The vertical axis corresponds to the initial parameter $z_1^0$ and
the plot shows, for each instant of time, whether the evolved state
is entangled (shaded area) or not (blank area). For instance, if the
initial state has $z_1^0
 = 2$, Simon's criterion will change from negative to positive in
 finite time; if one have $z_1^0 = 1.0$, the entanglement will vanish
 asymptotically. We will show an analytical relation between single-mode
 squeezing and ESD hereafter.


In the following we want to discuss the instant of time in which the
state becomes separable, or when occurs ESD, in the case of
\emph{zero temperature} and \emph{symmetrical states} ($z_{i}^0=0$,
$r_0>0$, $\nu_{i}^{0}$=$\nu_0$ and $\gamma_{i}=\gamma$,
$n_{i}^{B}=0$, for $i=1,2$). In this case the elements of the
covariance matrix, which do not vanish, depend on
\begin{align}
n& =n(t)=\frac{1}{2}e^{-2\gamma t}((2\nu_0 +1)\cosh (2r_0)-1)=n_{0}e^{-2\gamma
t}, \\
m& =m(t)=\frac{1}{2}e^{-2\gamma t}(2\nu_0 +1)\sinh (2r_0)=m_{0}e^{-2\gamma t}.
\end{align}%
In this case Simon $S$ can be factorized as $S=(m+n)(m+n+1)(m-n)(m-n-1)$.
Since the first two factors are clearly positive, we can see that $S$ is
negative if $n<m<n+1$. If this inequality is satisfied at the initial time,
$n_{0}<m_{0}<n_{0}+1$. Now, multiplying by $e^{-2\gamma t}$ we get
$n<m<n+e^{-2\gamma t}<n+1$. Thus, if the state is initially entangled,
the evolved state is also entangled for any finite time. Since Simon $S$
vanishes asymptotically  $\lim_{t\rightarrow \infty }S(t)=0$,
we see that either the entanglement decay is asymptotic, or the initial
state is already separable. A two-mode squeezed vacuum, $\nu_0=0$, is always
entangled; hence, it separates asymptotically.

Now we study $t_{ESD}$ for states containing single-mode squeezing, \emph{%
i.e.} $z_{1}=z_{2}$. Here one can find for $t_{ESD}$:
\begin{equation}
\mathrm{e}^{-2\gamma t_{ESD}}=\frac{2 \exp(r_0) \cosh 2z_0\sinh r_0-2\sinh ^{2}z_0}{%
\exp(2r_0)(\cosh 2r_0-\sinh 2z_0)},
\end{equation}%
where we consider $r_0>0$, and $z_1^0=z_2^0=z_0$.
In terms of the new variables $\eta =\exp(2r_0)$
and $\zeta = \exp(2z_0)$, the disentanglement time
\begin{equation*}
\mathrm{e}^{-2\gamma t_{ESD}}=\frac{\eta (1+\zeta ^{2}-2\eta \zeta )}{\eta
-\zeta -\eta ^{2}\zeta +\zeta ^{2}\eta },
\end{equation*}%
is much easier to analyze. This equation has a valid solution when the
right-hand side varies between 0 and 1, that is, when $\eta$ satisfies the
inequality $1\leq \eta \leq \frac{1}{2}\left( \zeta +\frac{1}{\zeta }\right)$.
Going back to the original variables we conclude that the initial state
separates at a finite time if $0<r_0<\frac{1}{2}\log \left( \cosh (2z_0)\right)$.
In this case, one can see clearly that the effect of single-mode squeezing
is crucial to determine when the entangle will vanish.

\subsection{Effects of the mixedness in the initial state}

\begin{figure}[htp]
\begin{center}
\vspace*{0cm} 
\includegraphics[bb = 18 9 341 266]{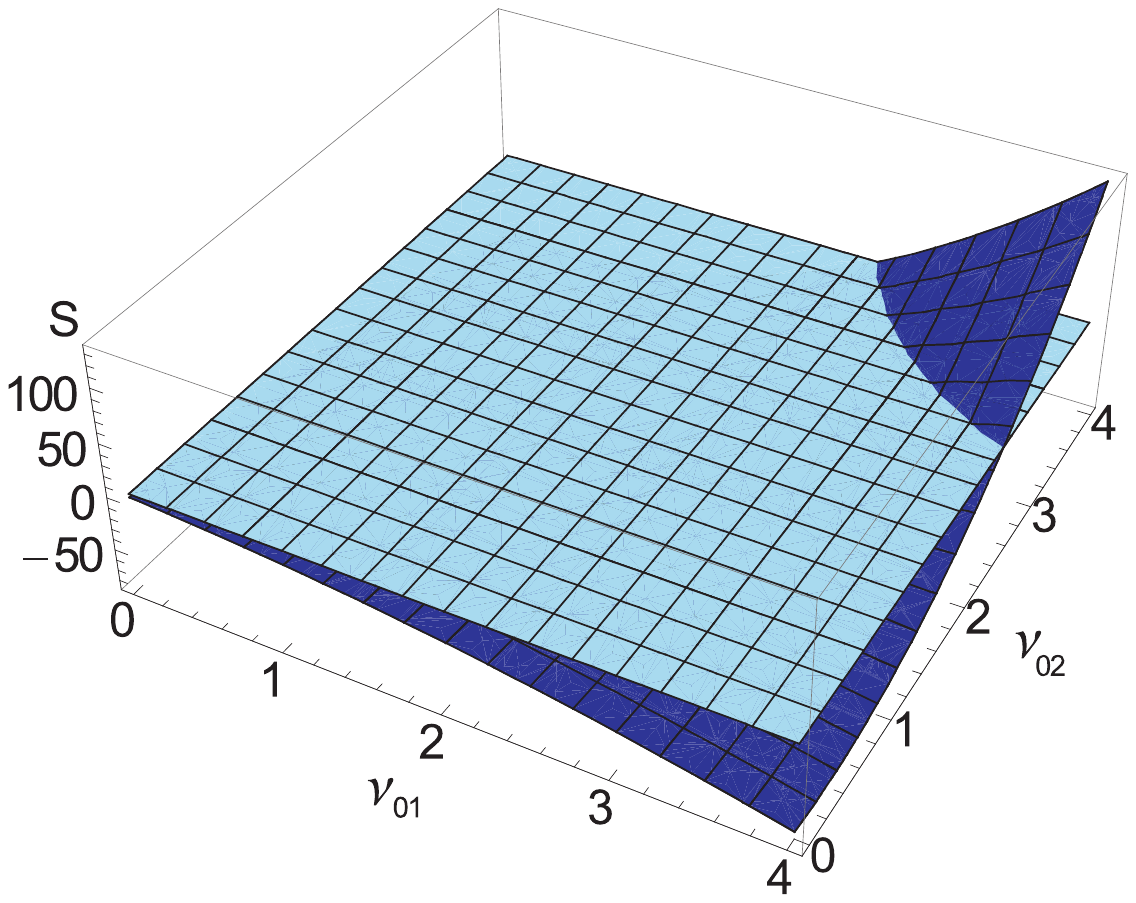}
\end{center}
\caption{Simon criterion for the two-mode Gaussian state, equation
\eqref{rhodoismodos}. Note that impurities values sufficiently high
can suppress the effect of entanglement given by the two-mode
squeezing parameter $r$. Parameter: $r_0 = 1$.} \label{figura9}
\end{figure}

Recently \cite{leocarol1, leocarol2} it has been shown that in the
case of single mode Gaussian states, there is an upper limit for the
degree of global purity (represented by $\nu_i^0$) of the state
above which no quantum properties are visible. We now show that an
analogous result holds for two mode Gaussian systems also. Following
the steps in ref. \cite{leocarol1, leocarol2} we can show that the
initial state is entangled if the following inequality is satisfied:
\begin{equation}
r_0> \frac{1}{4} \cosh^{-1} \left( \frac{(1 + \nu_2^0)^2 + 2 \nu_1^0 (1 +
\nu_2^0) (1 + 4 \nu_2^0) + (\nu_1^0)^2 (1 + 8 \nu_2^0 (1 + \nu_2^0))}{(1 +
\nu_1^0 + \nu_2^0)^2} \right).
\end{equation}
In Figure \ref{figura9} we show the entanglement in the initial state as
measured by $S$, with $r_0 = 1$. Notice that there is an upper limit on the
initial state impurity above which the state becomes separable.

\section{Conclusion}

In this paper we review some aspects of two-mode Gaussian states,
showing both the evolution of the state parameters (that entirely
characterize the state) under nonunitary evolution and how
entanglement of the state, characterized by the Simon criterion
\eqref{simonquantity}, depends on the initial state parameters. We
show that, even in completely symmetrical reservoirs and \emph{zero
temperature}, the entanglement can vanish in finite time, depending
on the single-mode squeezing of the state. We give a condition for
ESD relating single mode squeezing and two mode squeezing, that is
$0 < r_0 < \frac{1}{2} \log (\cosh (2 z_0))$. We analytically
present the time when occurs ESD for symmetrical states, evolving in
a reservoir with zero temperature. We also show that entanglement
can be \textquotedblleft suppressed\textquotedblright\ by the
initial mixedness of the modes, $\nu
_{i}^{0}$: the initial two mode squeezing has an upper limit as a function $%
\nu _{i}^{0}$, and for values above this limit the state is separable. This
can be helpful for experimental procedures, since any state will have some
minimal impurity.

\emph{Acknowledgements} - L.A.M. Souza thanks FAPEMIG for financial
support. L.A.M. Souza also thanks F. Toscano and F. Nic\'acio for
fruitful discussions and suggestion that they have made in the III
Workshop of Quantum Information (Paraty - $2011$).

\end{document}